\documentclass[12pt]{article}

\usepackage[numbers,round,sort&compress]{natbib}

\bibpunct{\textit{\textbf{(}}} {\textit{\textbf{)}}}{,}{n}{}{;}

\usepackage{amsmath,amssymb,xcolor,graphicx,wrapfig,paralist}
\usepackage[noend]{algpseudocode}
\usepackage{multirow}
\usepackage{booktabs}
\usepackage{proof}
\usepackage{tikz}
\usepackage[ruled, vlined]{algorithm2e}

\usetikzlibrary{shapes,positioning,arrows,calc,automata,matrix}
\tikzstyle{state}+=[minimum size = 8mm, inner sep=0,outer sep=1]
\tikzset{->,>=stealth'}

\definecolor{wwhite}{gray}{1}
\usepackage[T1]{fontenc}

\usepackage{tabularx}
\newcolumntype{L}{>{\raggedright\arraybackslash}p{1.6cm}}
\newcolumntype{C}{>{\centering\arraybackslash}p{1.6cm}}
\newcolumntype{R}[1]{>{\raggedleft\arraybackslash}p{#1}}
\allowdisplaybreaks

\newcommand{\thmhelperpre}[2]{\newcommand{\theoremlike}[1]{\par\medskip\penalty-250\refstepcounter{theorem}{\bfseries\noindent##1 \ref{#1}.}\itshape}\theoremlike{#2}}
\newcommand{\thmhelperpost}{\par\medskip%
 \renewcommand{\theoremlike}[1]{\par\medskip\penalty-250\refstepcounter{theorem}{\bfseries\noindent##1 \thesection .\thetheorem.}\itshape}%
}

\renewcommand{\vec}[1]{\mathbf{#1}}

\newcommand{\trerr}[1]{\xi}
\newcommand{\mperr}[1]{\zeta}

\def\p{p}

\newcommand{\srate}[1]{{c_{#1}}}

\newcommand{\conc}[1]{[{#1}]}

\newcommand{\xvec}{{\vec{x}}}

\newcommand{\spec}[1]{S_{#1}}
\def\specN{s}
\def\reacN{r}

\def\N{{\mathbb N}}

\newcommand\sprop[1]{\lambda_{#1}}

\newcommand\dprop[1]{\tilde{\lambda}_{#1}}

\def\volume{\Omega}

\newcommand{\EE}{\ensuremath{\mathsf{E}}}

\def\<{\langle}
\def\>{\rangle}
\def\~{\tilde}

\def\N{\mathbb N}

\def\R{\mathbb R}

\def\ra{\rightarrow}
\def\rA#1{\stackrel{{#1}}{\ra}}

\def\longrightharpoonup{\relbar\joinrel\rightharpoonup}
\def\longleftharpoondown{\leftharpoondown\joinrel\relbar}

\def\longrightleftharpoons{
  \mathop{
    \vcenter{
      \hbox{
    \ooalign{
      \raise1pt\hbox{$\longrightharpoonup\joinrel$}\crcr
      \lower1pt\hbox{$\longleftharpoondown\joinrel$}
    }
      }
    }
  }
}

\def\vunit{v}

\def\Xstov{\mathbf{X}}

\def\xdetabs{\tilde{\xdet}}
\def\xdetabsv{\tilde{\xdetv}}
\def\xstoabsv{\tilde{\xstov}}

\def\sp#1{\spec{#1}}
\def\xdet#1{[{#1}]}

\def\dt{\der t}

\def\consMatrix{P}
\def\changeMatrix{C}

\def\xdet{z} 
\def\xsto{x} 

\def\xdetv{\mathbf{z}}
\def\xstov{\mathbf{x}}
\def\ystov{\mathbf{y}}

\def\specN{n}
\def\reacN{r}

\def\N{{\mathbb N}}

\def\volume{{V}}

\def\drate{k}

\def\reac{\mathsf{r}}

\def\speciesSet{{\cal S}}

\def\reactionSet{\mathsf R} 

\def\frag{F}

\def\aut{Aut}

\def\der{\mathrm{d}}

\def\p{p}

\newcommand{\equref}[1]{(\ref{#1})}

\def\part{{\cal X}}

\def\ra{\rightarrow}
\def\N{{\mathbb{N}}}

\def\R{{\mathbb R}}

\def\ctmc{(\states,\cweight,\pizero)}

\def\cweight{w}

\def\pizero{\transient_0} 

\def\procX{X}

\def\cprocX{\{X_t\}}

\def\p{p} 

\def\transient{p}

\def\PP{\ensuremath{\mathsf{P}}}

\def\states{S}

\def\RBP{rule-based program}

\def\indic{\mathsf{1}}

\def\rA#1{\stackrel{{#1}}{\ra}}

\def\longrightharpoonup{\relbar\joinrel\rightharpoonup}
\def\longleftharpoondown{\leftharpoondown\joinrel\relbar}

\def\longrightleftharpoons{
  \mathop{
    \vcenter{
      \hbox{
    \ooalign{
      \raise1pt\hbox{$\longrightharpoonup\joinrel$}\crcr
      \lower1pt\hbox{$\longleftharpoondown\joinrel$}
    }
      }
    }
  }
}
\def\speciesSet{{\cal S}}

\def\aut{Aut}

\def\N{{\mathbb N}}

\def\<{\langle}
\def\>{\rangle}
\def\~{\tilde}

\def\N{\mathbb N}

\def\R{\mathbb R}

\def\ra{\rightarrow}
\def\rA#1{\stackrel{{#1}}{\ra}}

\def\longrightharpoonup{\relbar\joinrel\rightharpoonup}
\def\longleftharpoondown{\leftharpoondown\joinrel\relbar}

\def\longrightleftharpoons{
  \mathop{
    \vcenter{
      \hbox{
    \ooalign{
      \raise1pt\hbox{$\longrightharpoonup\joinrel$}\crcr
      \lower1pt\hbox{$\longleftharpoondown\joinrel$}
    }
      }
    }
  }
}

\def\change{\nu}

\usepackage{microtype}

\advance\textwidth2mm 
\advance\hoffset-1mm
\advance\textheight2mm
\advance\voffset-1mm

\usepackage[doublespacing]{setspace}

\begin{document}

\Large
\noindent \textbf{Markov chain aggregation and its application to rule-based modelling} \\

\large
\noindent Tatjana Petrov \\

\normalsize
\noindent Department of Computer and Information Sciences, University of Konstanz, Konstanz, Germany \\

\vspace{0.5in}

\noindent \textbf{Running head}: Markov chain aggregation \\

\noindent \textbf{Corresponding author}: \\ Tatjana Petrov \\ Department of Computer and Information Sciences \\ University of Konstanz \\ Konstanz, Germany \\ Email: tatjana.petrov@uni-konstanz.de \\
\emph{This article has been prepared for Springer book series `Methods in Molecular Biology', volume title `Modeling Biomolecular Site Dynamics'.
Tatjana Petrov's research was supported by the Ministry of Science, Research and the Arts of the state of Baden-W\"{u}rttemberg.}

\clearpage

\section*{Summary}

Rule-based modelling allows to represent molecular interactions in a compact and natural way. The underlying molecular dynamics, by the laws of stochastic chemical kinetics, behaves as a continuous-time Markov chain. However, this Markov chain enumerates all possible reaction mixtures, rendering the analysis of the chain computationally demanding and often prohibitive in practice. We here describe how it is possible to efficiently find a smaller, aggregate chain, which preserves certain properties of the original one. Formal methods and lumpability notions are used to define algorithms for 
automated and efficient construction of such smaller chains (without ever constructing the original ones). We here illustrate the method on an example and we discuss the applicability of the method in the context of modelling large signalling pathways. \\

\noindent
\textbf{Key words}: Markov chain aggregation, lumpability, bisimulation, rule-based modelling

\clearpage

\section{Introduction}

After gaining new possibilities for experimenting, by the development of fluorescent biomarkers for proteins, detection of RNA and interactions, microfluidic technology, high-resolution imaging, biology seeks appropriate mechanistic explanations of the obtained measurements. Systems and synthetic biology aim at systemic, quantitative understanding of molecular processes, for both explanatory (scientific) and practical (engineering) purposes.

\subsection{General background}

The ground model of biochemical network dynamics is given by {stochastic chemical kinetics}:
under certain simplifying assumptions, a low-level description of the dynamics of a biochemical network is captured by a continuous-time Markov process (CTMC), in which one state corresponds to one reaction mixture, encoded as a multi-set of chemical species. 
For example, a state can be $x=[2S_1,3S_2,5S_3]$, where $S_1$, $S_2$, $S_3$ are chemical species. Then, a reaction, for example, $S_1,S_2\rightarrow S_3$ takes the system from the state $x$ to the state $x'=[S_1,2S_2,6S_3]$, at a stochastic rate which is defined in the physical chemistry domain.
A system becomes huge both as the number of reactions and the number of species increase.
Such species-centered models have yet another source of complexity: if proteins $A$ and $B$ have respectively $n$ and $m$ domains which can all receive a phosphorylation signal, then there is $2^n+2^m+2^{n-1}2^{m-1}$ different molecular species formed only by these two molecules.  
For example, one model of the early signaling events in the epidermal growth factor receptor (EGFR) signaling network, which accounts for only $8$ different proteins, gives rise to 2,748 different molecular species \cite{blinov2006network}. 
To this end, modeling with traditional chemical kinetics faces fundamental limitations, related to the question of how biochemical events are represented. 

One way of dealing with the complexity of cellular signalling  is using \emph{formal models}, which allow to execute models from a collection of machine-readable instructions (Figure~1). 
One approach in this direction are \emph{rule-based models} (implemented in either Kappa \cite{danos2004formal} or BioNetGen \cite{RB16} formats), proposed for modelling signalling pathways in cells: they are designed to capture low-level molecular interactions. Importantly, they support expressing a state-change by testing only states of proteins' domains, instead of the full molecular complexes. 
More precisely, take a protein $A$ with domains $s$ and $t$, such that each of them could have received phosphorylation or not. 
Then, a spontaneous phosphorylation of the site $s$ is captured by a rule
$A(s\sim u)\rightarrow A(s\sim p). $
So, the syntax of the language allows to express naturally `protein $A$ whose state $s$ is unphosphorylated'.
Such syntax clearly reflects that the logic behind the design of rule-based models takes  parts of species, \emph{patterns}, as main entities of observation (information carriers). Indeed, it was shown that a protein-centric representation naturally benefits in more efficient simulations \cite{danos:aplas07}. 
However, for precise analysis of stochastic behaviours the full underlying CTMC must be considered, that is, the enumeration of all reaction mixtures cannot be avoided. 

A small number of rules can generate a system of astronomical state space \cite{RB4,sorger06}, rendering the expansion to the species-based description often infeasible even to write down.
However, since the huge state space emerges from a small number of rules operating over patterns, 
there is hope to capture the dynamics of a rule-set compactly, as a function of patterns, which are much fewer than the full molecular species. 
For that reason, we try to detect those patterns, called \emph{fragments}, which can faithfully describe the dynamics of a rule-set.
The term `fragment' is chosen in the sense that it is syntactically represented as \emph{a fragment of a full species}.

We here illustrate over an example the method for obtaining mechanistic predictions about stochastic rule-based models at a level of patterns (fragments), while using the theory of Markov chain aggregation, based on our works \cite{reconstruction,approx,journal,lumpability}. 
The method is automatic, so it is not a heuristic solution which works for a certain case study, but a general method which can be used for any rule-based model.
The properties of the reduced model are ensured by establishing a lumpability (bisimulation) relation between the original and reduced model. 

We introduce stochastic chemical kinetics and rule-based models in the following section. 
In the Methods section, we illustrate exact stochastic fragment-based reduction for a particular example. 
Then we demonstrate the method applied to a case study of EGF/insulin crosstalk and we conclude with a discussion and suggestions for future work.

\subsection{Mathematical background}

Population models are widely used in modelling interactions among a set of individuals, distinguishable only by the class of species they belong to. 
Population models can be represented in terms of reactions of the form
$
{A+2B}
\rA{k} 
{C}
$, 
where $A$ and $B$ are reactant species, $C$ is the product species, and $k$ is a parameter that characterizes the {rate} or a speed at which the change occurs. 

\newcommand{\consv}[1]{{\boldsymbol a}_{#1}}
\newcommand{\gainv}[1]{{\boldsymbol \nu}{#1}}
\newcommand{\changev}{\boldsymbol{\nu}}
\newcommand{\cons}[1]{{a}_{#1}}
\newcommand{\gain}[1]{{b}{#1}}
\newcommand\prdn[1]{a'_{#1}}

Let us formally define a reaction system. A \textit{reaction system} is a pair $(\speciesSet,\reactionSet)$, such that
\begin{enumerate}
\item $\speciesSet=\{\spec{1},\spec{2},\ldots,\spec{\specN}\}$ is a finite set of species,
\item 
$\cons{1j}\spec{1},\ldots,\cons{\specN j}\spec{\specN} \rA{\drate_{j}}\prdn{1j}\spec{1},\ldots,\prdn{\specN j}\spec{\specN}, \hbox{ such that $\prdn{ij}=\cons{ij}+\change_{ij}$}.$
The vectors $\consv{j}$ and $\consv{j}'$ are often called respectively the \emph{consumption} and \emph{production} vectors due to reaction $\reac_j$, $k_j$ is the \emph{kinetic rate} of reaction $\reac_j$ and $\changev_{j}=\consv{j}'-\consv{j}$ is called the \emph{change} vector.
\end{enumerate}

A model of population dynamics can be
(i) discrete or continuous, depending on whether the population quantity is modeled as a discrete or a continuous value, 
 and 
(ii) {deterministic} or {stochastic}, depending on whether the 
output trajectory is fully determined by the initial state (deterministic), or if different trajectories can emerge, each associated with a certain probability (stochastic).

Classical chemical kinetics handles ensembles of molecules with large number of particles, $10^{20}$ and more. The chemist uses concentrations rather than particle numbers, $\conc{N}=N/N_A\cdot \volume$, where $N_A=6.23\cdot10^{23}$ mol${}^{-1}$ is the Avogadro's number and $\volume$ is the volume (in dm${}^3$).
When the pressure and temperature are constant, the following continuous, deterministic model is appropriate.

Let $(\speciesSet,\reactionSet)$ be a reaction system, and $\xdetv_0=(\xdet_1,...,\xdet_{\specN})\in \R^{\specN}$ an initial state of the system.
Then, the \emph{continuous, deterministic model} is the solution of the set of $\specN$ coupled differential equations given by 
\begin{align}\label{eq:CCK}
\frac{\der}{\der t} \xdet_i(t) = \sum_{j=1}^{\reacN} \change_{ij}\dprop{j}(\xdetv(t)), \hbox{ for $i=1,2,\ldots,\specN$}, 
\end{align}
satisfying the initial condition $\xdetv_0$.
The family of functions $\{\dprop{j}:\R^{\specN}\ra\R\mid {j=1,\ldots,\reacN}\}$, called also \emph{deterministic reaction rates} is defined by
\begin{align} \label{eq:fpre}
\dprop{j}(\xdetv)= \drate_j\prod_{i=1}^{\specN} {z_i}^{\cons{ij}}.
\end{align}
The fact that the speed of a chemical reaction is proportional to the quantity of the reacting substances
is known as the \emph{kinetic law of mass action}.

It was shown that stochastic effects generate phenotypic heterogeneity in cell behavior and that cells can functionally exploit variability for increased fitness (\cite{noisy_bussiness} is an early review on the subject).
As many genes, RNAs and proteins are present in low copy numbers, 
 deterministic models are insufficiently informative or even wrong.
For example, for a simple  birth-death model 
$\emptyset\rA{k_1} S_1,\;\;S_1\rA{k_2} \emptyset$,
the deterministic solution $z(t)=z(0)e^{t(k_1-k_2)}$ 
is interpreted as the mean population of species $S_1$ through time. 
Any additional experimental observation, such as the degree of deviation around the average value, or
 the probability of extinction of the species at a given time cannot be deduced.
In more complex examples, observing that the population exhibits bimodal response cannot be made unless a stochastic model is employed. 

A discrete, stochastic model of a biochemical reaction system, 
reacting in a well-stirred mixture of volume $\volume$ and in thermal equilibrium is defined below.
This definition can be derived from the fundamental premise of stochastic chemical kinetics \cite{gillespie_92markov}.

Let $(\speciesSet,\reactionSet)$ be a reaction system, and $\xstov_0=(\xsto_1,...,\xsto_{\specN})\in \N^{\specN}$ an initial state of the system.
Then, the \emph{discrete, stochastic model} is a continuous-time Markov chain (CTMC) $\{\procX_t\}$
over the set of states $\states=
\{\xstov\mid \hbox{$\xvec$ is reachable from $\xstov_0$ in $\reactionSet$}\}$,
initial probability $\pizero(\xstov_0)=1$,
with the generator matrix defined by
$\cweight(\xstov,\ystov) = \sum \{\sprop{j}(\xstov)\indic_{\ystov=\xstov+\changev_j}\mid {j=1,\ldots,\reacN}\}$.
The family of functions $\{\sprop{j}:\R^{\specN}\ra\R\mid {j=1,\ldots,\reacN}\}$, called also \emph{stochastic reaction rates}, is defined by
\begin{align} \label{eq:fpre}
\sprop{j}(\xstov)= 
\srate{j}\prod_{i=1}^{\specN}{x_i\choose\cons{ij}}
\end{align}
The binomial coefficient ${x_i\choose\cons{ij}}$ represents the probability of choosing $\cons{ij}$ molecules of species $S_i$ out of $x_i$ available ones.

Using the vector notation $\Xstov_t$ for the marginal of process $\cprocX$ at time $t$,  
we are typically interested in the transient probability distribution of $\cprocX$, 
which can be obtained by solving the \emph{chemical master equation} (CME): for
$p^{(t)}(\xstov)=\PP(\Xstov_t=\xstov)$,
the CME for state $\xstov\in\N^{\specN}$ is
\begin{align}
\frac{\der}{\der t} p^{(t)}(\xstov) = 
\sum_{j=1, \xstov-\changev_{j}\in\speciesSet}^{\reacN} \sprop{j}(\xstov-\changev_{j}) p^{(t)}(\xstov-\changev_{j}) -
\sum_{j=1}^{\reacN} \sprop{j}(\xstov)p^{(t)}({\xstov}).
\end{align}
The solution may be obtained by solving the system of equations, but due to its high dimensionality, it is more often statistically estimated by simulating the traces of $\cprocX$, via a procedure known as the stochastic simulation algorithm (SSA) in the chemical literature \cite{Gillespie}.

Notice that the CME implies that the expectation of the marginal distribution of $\cprocX$ satisfies the equations
$\frac{\der}{\der t} \EE(\Xstov_t) = \sum_{j=1}^{\reacN} \changev_{j}\EE(\sprop{j}(\Xstov_t))$.
It is worth noting that, upon scaling the rate constants, the equations for $\EE(\Xstov_t)$ are equivalent to \equref{eq:CCK} only if 
all rate functions are linear, that is, when all reactions are unimolecular. 

We mentioned above the existence of both a reaction rate constant $k_j$ and a stochastic rate constant $c_j$. These deterministic  and stochastic rate constants are not equivalent. When switching 
between the stochastic and deterministic model, a conversion of rates must be performed. 
In particular, the stochastic rate constant depends on the volume and the molecularity of a reaction. 
In general, the conversion is such that the stochastic rate function applied to a state $\xvec\in\N^{\specN}$
for a reaction $\reac_j$, and the deterministic law of its conversion to a volume unit---$\xvec^{\volume}\in\R^{\specN}$---will relate as 
$\dprop{j}({\xvec}{\volume}^{-1}) = \sprop{j}(\xvec)\volume^{-1}$.
The careful study of the above conversions is outlined in \cite{gillespie_92markov}. Intuitively,  
 observe that, as unimolecular reactions represent a spontaneous conversion of a molecule, they should not be volume dependent. 
In bimolecular reactions, the stochastic rate $c_j$ will be proportional to $1/\volume$, reflecting that two molecules have a harder time finding each other within a larger volume.  

Even though deterministic models historically appeared first, they represent a particular approximation of the stochastic model, in a limit in which the reactant populations $x_i$ and the system volume $\volume$ all become infinitely large, but in such a way that the reactant concentrations $x_i/\volume$ stay fixed \cite{Anderson2010}.

A rule-based language can be viewed as a form of site-graph-rewrite grammar, designed
for modeling low-level bio-molecular interactions. 
A rule-based model can be understood as a compact, symbolic 
encoding of a set of biochemical reactions. 
A simple rule-based model is sketched in Figure 2.
Informally, an agent of type $B$ can form a bond with either an agent of type $A$ or an agent of type $C$, via specific (typed) site variables ($a$, $b$ or $c$).
A transition can be triggered upon local tests on an agent's interface; omitting the site $c$ of agent $B$ in rule $R_1$ (or $R_1^-$) means that the conformation of site $c$ is irrelevant for executing rule $R_1$ (or $R_1^-$) (sometimes referred to as the 
 \emph{don't care, don't write} agreement). Typically, agent types encode proteins and site types encode respective protein domains.
The executions of rule-based models---programs written in a rule-based language---are defined according to the principles of stochastic chemical kinetics, established in the physical chemistry and molecular physics domain. 
We illustrate both the syntax and semantics of rule-based models for a simple example, Example 1 (described immediately below).
The variants of operational semantics can be found in \cite{lumpability} and references therein.

\noindent \textbf{Example 1}: a simple model for interactions of a scaffold protein. 
Scaffold protein $B$ recruits independently the proteins $A$ and $C$.
These assumptions are captured by a set of rules, $ \{R_1,R_2,R_1^-,R_2^-\}$ depicted in Figure 2.
Adding the rules $R_3$ and $R_4$ accelerates the unbinding, whenever the bond is within a trimer complex (that is, the bonds are made less stable when within a trimer).

The corresponding reaction system is $(\speciesSet,\reactionSet)$, where
$
\speciesSet = \{\sp{A},\sp{B},\sp{C},\sp{AB},\sp{BC},\sp{ABC}\}
$
and 
$
\reactionSet = 
\{
\reac_{A.B},\reac_{B.C},\reac_{A.BC},\reac_{AB.C},\reac_{A..B},\reac_{B..C},\reac_{A..BC},\reac_{AB..C}
\}
$, defined by
\begin{align*}
\reac_{A.B}:  \;\;\;\;&
\sp{A},\sp{B} \rA{\drate_{1}} \sp{AB}  \;\;\;\;\\
\reac_{A.BC}:  \;\;\;\;&
 \sp{A},\sp{BC} \rA{\drate_{1}} \sp{ABC}\;\;\;\;\\
\reac_{B.C}:  \;\;\;\;&
\sp{B},\sp{C} \rA{\drate_{2}}\sp{BC} \;\;\;\;\\
\reac_{AB.C}: \;\;\;\;&
\sp{AB},\sp{C} \rA{\drate_{2}}\sp{ABC}\\
\reac_{A..B}:  \;\;\;\;&
\sp{AB} \rA{\drate_{1^-}}  \sp{A},\sp{B}\;\;\;\;\\
\reac_{A..BC}:  \;\;\;\;&
\sp{ABC}\rA{\drate_{1^-}}\sp{A},\sp{BC}  \;\;\;\;\\
\reac_{B..C}:  \;\;\;\;&
\sp{BC} \rA{\drate_{2^-}}\sp{B},\sp{C}\;\;\;\;\\
\reac_{AB..C}:  \;\;\;\;&
\sp{ABC}\rA{\drate_{2^-}}\sp{AB},\sp{C}. 
\end{align*}

The consumption vectors and change vectors are the column vectors of matrices $\consMatrix$ and $\changeMatrix$:
\[ \consMatrix = 
\left( 
\begin{array}{ccccccccc}
1 & 1 & 0 & 0 & 0 & 0& 0 & 0 \\
1 & 0 & 1 & 0 & 0 & 0& 0 & 0 \\
0 & 0 & 1 & 1 & 0 & 0& 0 & 0 \\
0 & 0 & 0 & 1 & 1 & 0& 0 & 0 \\
0 & 1 & 0 & 0 & 0 & 0& 1 & 0 \\
0 & 0 & 0 & 0 & 0 & 1& 0 & 1 
\end{array} \right)\hbox{ and }
\changeMatrix=
\left(
\begin{array}{ccccccccc}
-1 & -1 & 0 & 0 & 1 & 1& 0 & 0 \\
-1 & 0 & -1 & 0 & 1 & 0& 1 & 0 \\
0 & 0 & -1 & -1 & 0 & 0& 1 & 1 \\
1 & 0 & 0 & -1 & -1 & 0& 0 & 1 \\
0 & -1 & 1 & 0 & 0 & 1 & -1 & 0 \\
0 & 1 & 0 & 1 & 0 & -1& 0 & -1 
\end{array} \right),
\] 
where, according to mass-action kinetics, the rate function has the following form:
\begin{align*}
\dprop{}(\xdetv) = (\drate_{1}\xdet_{A}\xdet_{B}, 
\drate_{1}\xdet_{A}\xdet_{BC},
\drate_{2}\xdet_{B}\xdet_{C},
\drate_{2}\xdet_{AB}\xdet_{C},
\drate_{1^-}\xdet_{AB}, 
\drate_{1^-}\xdet_{ABC},
\drate_{2^-}\xdet_{BC},
\drate_{2^-}\xdet_{ABC}
).
\end{align*}

\noindent
\textbf{A deterministic model for the system of Example 1.}
\label{sec:detMod}
Denote by $\xdetv\in\R^6$ the vector of concentrations of species from $\speciesSet$.
For keeping transparency, let $\xdet_A$ denote the concentration of species $A$,  $\xdet_B$ the concentration of species $B$ etc.
The continuous, deterministic model is given by the set of ordinary differential equations:
\begin{align*}
{\der\xdet_{A}} \over {\dt} &=-\xdet_{A}\xdet_{B}\drate_{1}-\xdet_{A}\xdet_{BC}\drate_{1}+\xdet_{AB}\drate_{1^-}+\xdet_{ABC}\drate_{1^-}\\
{\der\xdet_{B}} \over {\dt} & =-\xdet_{A}\xdet_{B}\drate_{1}-\xdet_{B}\xdet_{C}\drate_{2}+\xdet_{AB}\drate_{1^-}+\xdet_{BC}\drate_{2^-}\\
{\der\xdet_{C}} \over {\dt} & =-\xdet_{B}\xdet_{C}\drate_{2}-\xdet_{AB}\xdet_{C}\drate_{2}+\xdet_{BC}\drate_{2^-}+\xdet_{ABC}\drate_{2^-}\\
{\der\xdet_{AB}} \over {\dt} & =-\xdet_{AB}\xdet_{C}\drate_{2}-\xdet_{AB}\drate_{1^-}+\xdet_{A}\xdet_{B}\drate_{1}+\xdet_{ABC}\drate_{2}\\
{\der\xdet_{BC}} \over {\dt} & =\xdet_{B}\xdet_{C}\drate_{2}-\xdet_{BC}\drate_{2^-}+\xdet_{B}\xdet_{C}\drate_{2}+\xdet_{ABC}\drate_{1^-}\\
{\der\xdet_{ABC}} \over {\dt} & =-\xdet_{ABC}\drate_{1^-}-\xdet_{ABC}\drate_{2^-}+\xdet_{A}\xdet_{BC}\drate_{1}+\xdet_{AB}\xdet_{C}\drate_{2}.
\end{align*}

\noindent
\textbf{A stochastic model for the system of Example 1.}
Assume that there are initially three copies of agent $B$, one copy of agent $A$ and one copy of agent $C$, which is represented by a population state $\xstov_0=(1,3,1,0,0,0)$.
For transparency, we will represent states in form of multi-sets - for example,
$\xstov_0\equiv \{A,3B,C\}$.
The stochastic model is a CTMC $\cprocX$ with a Markov graph,$\ctmc$, such that
$\pizero(\xstov_0)=1$,
$\states = \{\xstov_0,\xstov_1,\xstov_2,\xstov_3,\xstov_4\}$, and the weights are as depicted in Figure 3.

Denoting by $p^{(t)}(\xstov)=\PP(X_t=\xstov)$, the CME 
is represented by the following system of equations (the superscript $^{(t)}$ is omitted):
\begin{align*}
{\der \p(\xstov_0)} \over {\dt} &= \srate{1^-}\p(\xstov_1)+\srate{1^-}\p(\xstov_2)-\p(\xstov_0)(3\srate{1}+3\srate{2})\\
{\der \p(\xstov_1)} \over {\dt} &= 3\srate{1}\p(\xstov_0)+\srate{2^-}\p(\xstov_3)+\srate{2^-}\p(\xstov_4)-\p(\xstov_1)(\srate{1^-}+2\srate{2}+\srate{2})\\
{\der \p(\xstov_2)} \over {\dt} &= 3\srate{2}\p(\xstov_0)+\srate{1^-}\p(\xstov_3)+\srate{1^-}\p(\xstov_4)-\p(\xstov_2)(\srate{1^-}+2\srate{1}+\srate{1})\\
{\der \p(\xstov_3)} \over {\dt} &= 2\srate{2}\p(\xstov_1)+2\srate{1}\p(\xstov_2)-\p(\xstov_3)(\srate{2^-}+\srate{1^-})\\
{\der \p(\xstov_4)} \over {\dt} &= \srate{2}\p(\xstov_1)+\srate{1}\p(\xstov_2)-\p(\xstov_4)(\srate{2^-}+\srate{1^-}).
\end{align*}

In Figure 4a, we show the solution of the model in the deterministic limit, and one trajectory of a stochastic model scaled with the volume, $\Xstov^{\volume}$.
In Figure 4b, we illustrate that, due to bimolecular reactions, the mean population size does not coincide with the solution in the deterministic limit.
The used values of rate constants are not inspired from real data.
A volume unit is denoted by $\vunit$.
To compare the deterministic and stochastic models, we assume that the volume scales with the total molecule number, more precisely, that one volume unit corresponds to five molecules.
Therefore, for the initial state for the stochastic model $\xstov(0)=(20,60,20,0,0,0)$ (molecules),
 the volume of $\volume=100$ molecules takes $20$ units, i.e., $\volume=20\vunit$.

\section{Materials}

Kappa-formatted definitions of the models discussed in this chapter are provided as Supplementary Materials.

\section{Methods} 

We are now ready to discuss \emph{fragment-based} reductions for stochastic rule-based models, and the role of Markov chain aggregation in these reductions. 

\subsection{Deterministic fragments}

We first illustrate the notion of fragments for fragments that preserve deterministic semantics. Let us provide a definition of \textit{deterministic fragments} for the system of Example 1. 
We consider a projection from a system state $\xdetv(t)$ to a state $\xdetabsv(t)$ with three components
$\{\xdetabs_{A},\xdetabs_{B?},\xdetabs_{AB?}\}$, such that 
\begin{align}
\label{eq:soundEx}
\xdetabs_{A}(t)&=\xdet_{A}(t)\\ \nonumber
\xdetabs_{B?}(t)&=\xdet_{B}(t)+\xdet_{BC}(t)\\ \nonumber
\xdetabs_{AB?}(t)&=\xdet_{AB}(t)+\xdet_{ABC}(t)\\ \nonumber
\end{align}

Looking back at the system of ODE's,
since differentiation is a linear operator,
the derivatives of the new variables compute to 
\begin{align}\label{eq:ODEproj}
{\der\xdetabs_{A}} \over {\dt} &=-\xdetabs_{A}\xdetabs_{B?}\drate_{1}+\xdetabs_{AB?}\drate_{1^-}\\\nonumber
{\der\xdetabs_{B?}} \over {\dt} & = 
-\drate_{1}\xdetabs_{A}\xdetabs_{B?}+\drate_{1^-}\xdetabs_{AB?}
\\
{\der\xdetabs_{AB?}} \over {\dt} & =\nonumber
\drate_{2}\xdetabs_{A}\xdetabs_{B?}-\drate_{1^-}\xdetabs_{AB?}.
\end{align}
The system \equref{eq:ODEproj} operates only over
 the variables $\{\xdetabs_{A},\xdetabs_{B?},\xdetabs_{AB?}\}$, that is, it self-consistently describes their dynamics. 
By solving the smaller system \equref{eq:ODEproj}, the full dynamics of the concrete system is not known, but meaningful information about the original system is obtained.

The system \equref{eq:ODEproj} is exactly the deterministic semantics of a reaction model 
\begin{align}
\label{eq:reduced}
F_A, F_{B?}&\rA{\drate_1} F_{AB?} \\
F_{AB?}&\rA{\drate_{1^-}} F_A, F_{B?}\nonumber
\end{align}
operating over three `abstract species',
denoted by $\frag_{A}$, $\frag_{B?}$ and $\frag_{AB?}$.
These `abstract species' are called {fragments}. 
In particular, notice that, for example, the contribution of fragment $F_{B?}$ with respect to rule $R_2$ is zero.
This is because $F_{B}$ is consumed at rate $\drate_2\xdet_B\xdet_C$, while $F_{BC}$ gets produced at the same rate.
The two terms cancel out, and  we say that rule $R_2$ is \emph{silent} with respect to $F_{B?}$. 

Fragment-based reduction schemes aim to immediately derive the system \eqref{eq:reduced}, in contrast to first expanding the equivalent species-based description, and then detecting symmetries in the equations.
To this end, this method is different from other principled model simplification techniques, based on, for example,
separating time-scales  \cite{rao2003stochastic,kang2013separation,ovidiu} or exploiting conservation laws \cite{conzelmann2008,borisov:08}.
In fragment-based reductions, the species-based system is 
considered only for the purpose of proving the relation between the reduced and the original model.
Once a fragment-based rule set is obtained, it is amenable to any further analysis.

These reductions have been termed \emph{fragment-based} by Feret and co-workers, who used them for automatically reducing the deterministic semantics of rule-based models \cite{pnas2009}. 
Below, we will consider the same example (Example 1) to illustrate the fragment-based technique for reducing stochastic semantics of rule-based models, that is, characterizing the \emph{stochastic fragments} and computing their dynamics. 

\subsection{Stochastic fragments}

In Figure 3a, the stochastic model for initially one copy of free $\spec{A}$, one copy of free $\spec{C}$ and three copies of free $\spec{B}$ is represented. 
The description in terms of fragments $\{F_{A}, F_{B?}, F_{AB?}, F_C,\frag_{?BC}\}$ means that states $\xstov_3$ and $\xstov_4$ are indistinguishable.
Let $\xstoabsv_{34} := \xstov_3 + \xstov_4$.
Then, we can compute the evolution of the fragment-based states: 
\begin{align*}
{\der \p(\xstoabsv_{34})\over {\dt}} 
& = {\der \p(\xstov_3) \over {\dt} }+ {\der \p(\xstov_4)\over {\dt}}\\
& = 3\srate{2}\p(\xstov_1)+3\srate{1}\p(\xstov_2)-(\srate{2^-}+\srate{1^-})(\p(\xstov_3)+\p(\xstov_4))\\
& = 3\srate{2}\p(\xstov_1)+3\srate{1}\p(\xstov_2)-(\srate{2^-}+\srate{1^-})\p(\xstoabsv_{34})\\
{\der \p(\xstov_1)} \over {\dt} &= 3\srate{1}\p(\xstov_0)+\srate{2^-}\p(\xstov_3)+\srate{2^-}\p(\xstov_4)-\p(\xstov_1)(\srate{1^-}+2\srate{2}+\srate{2})\\
&=3\srate{1}\p(\xstov_0)+\srate{2^-}\p(\xstoabsv_{34})-\p(\xstov_1)(\srate{1^-}+2\srate{2}+\srate{2})\\
{\der \p(\xstov_2) \over {\dt} }&=  3\srate{2}\p(\xstov_0)+\srate{1^-}\p(\xstov_3)+\srate{1^-}\p(\xstov_4)-\p(\xstov_2)(\srate{1^-}+2\srate{1}+\srate{1})\\
&=
3\srate{2}\p(\xstov_0)+\srate{1^-}\p(\xstoabsv_{34})-\p(\xstov_2)(\srate{1^-}+2\srate{1}+\srate{1}),
\end{align*}
Because the above set of equations is self-consistent,
the CTMC in Figure 3b can be used to compute the transient distribution of the lumped process: the probability of being in a state $\xstoabsv_{34}$ is the sum of probabilities of being in states $\xstov_3$ and $\xstov_4$. This property of a chain with respect to a given partition of states is called \emph{lumpability} (\textit{see} \textbf{Note 1}).

It turns out that from the lumped process we can also recompute the trace distribution of the original process, a property which is termed \emph{invertability} (of the aggregate chain with respect to the given partition and a distribution): 
the conditional probability of being in a state $\xstov_3$ or $\xstov_4$ can be recovered from that of $\xstoabsv_{34}$. 
In particular, the theory confirms that the ratio between the probability $\p^{(t)}(\xstov_3)$ and $\p^{(t)}(\xstov_4)$ can be reconstructed as the ratio of automorphisms of site-graphs which represent the states $\xstov_3$ and $\xstov_4$ respectively \cite{reconstruction,tanja_thesis}:
\begin{align}\label{eq:tako}
\frac{\p^{(t)}(\xstov_3)}{\p^{(t)}(\xstov_4)}=\frac{|\aut(\{\spec{AB},\spec{B},\spec{BC}\})|}{|\aut(\{\spec{ABC},2\spec{B}\})|}=\frac{2}{1}.
\end{align}
To check that \eqref{eq:tako} holds,
let $\Delta(t) := \frac{1}{2}\p^{(t)}(\xstov_3)-\p^{(t)}(\xstov_4)$.
Then,  
\begin{align*}
{\der \Delta(t) \over {\dt} }=-(\srate{2^-}+\srate{1^-})\Delta(t)
\end{align*}
 has a unique solution $\Delta(t)=\Delta(0)e^{-(\srate{2^-}+\srate{1^-})t}$, meaning that the probability of being in state $\xstov_3$ converges to being exactly two times larger than the probability of being in state $\xstov_4$, and, combined with the self-consistency derivation, it follows that $\p^{(t)}(\xstov_3)=\frac{2}{3}\p^{(t)}(\xstoabsv_{34})$.
If $\Delta(0)=0$, the ratio between probabilities will always hold, and otherwise it will be the case asymptotically. 

Importantly, the conclusions drawn above are not valid in a case where, for example, the rate of unbinding $\spec{ABC}$ is stronger than the rate of unbinding $\spec{AB}$ or $\spec{BC}$ separately. In this case, it would not be possible to write the equation for ${\der \p(\xstov_1)} \over {\dt}$ and for ${\der \p(\xstov_2)} \over {\dt}$ as a function of $\p(\xstoabsv_{34})$.
In this case, the proposed fragmentation is not \emph{expressive} enough, since it cannot express a quantity which is necessary for the correct description of the fragment dynamics.
Consequently, any proposed reduction with the same choice of fragments will only be approximate. 

\subsection{Fragmentation algorithm}

The goal of exact fragment-based reductions of stochastic rule-based models is to generalize the made observations, so that the presented reduction can be detected and performed on any \RBP. 
The detection of fragments involves characterizing the states of the CTMC that can be lumped while preserving the lumpability (and potentially invertability) relation.
In the above example, to claim the properties it suffices to establish that the CTMC in Figure 5a is \emph{lumpable} with respect to the partition which merges the states $\xstov_3$ and $\xstov_4$, or, 
equivalently, that the states $\xstov_3$ and $\xstov_4$ are \emph{backward bisimilar} \cite{mathpaper}.
Ensuring these relations hold boils down to detecting groups of sites that a rule-set must simultaneously `know' in order to execute the rules without error. 
For example, executing a rule $R_3$ in Example 1 demands determining whether the species $S_{ABC}$ embeds into the current reaction mixture, implying that the correlation between connectivity of sites $a$ and $c$ on node type $B$ must be maintained.

The sketch of the general fragmentation algorithm is shown in Figure 6.
The input to the fragmentation process is
(i) the set of observable species, patterns or their combination within a reaction soup (for example, we may be interested in the average copy number of $S_A$ and $S_C$, or the probability of being in the state with 100
patterns $F_{AB?}$ and 100 patterns $F_{?BC}$); and
(ii) the rule-set.
The fragments are chosen so that the dynamics of the observables can be correctly and self-consistently computed from the fragment-based description.
The formal introduction and proofs of the mentioned concepts can be found in \cite{lumpability,mathpaper,tanja_thesis}. We note that the goal of the fragmentation procedure discussed here is efficiency (\textit{see} \textbf{Note 2}).

\subsection{Application to a model for EGF/insulin receptor signaling crosstalk}
\label{ex:egfr}

The method was applied on a model of a crosstalk between the epidermal growth factor (EGF) receptor (EGFR) and the insulin (Ins) receptor (IR) pathway.
EGFR is present on the cell surface and is activated by binding of its specific ligands, such as EGF.
Upon ligand binding, EGFR initiates a signaling cascade entailing a variety of biochemical changes that influence processes such as cell growth, proliferation, and differentiation.
A huge number of feasible multi-protein species can be formed in this signaling pathway \cite{blinov2006network}. 
For example, in the model described in \cite{lics2010}, the number of reachable complexes is estimated to be $\approx 10^{20}$.
We focused on a model of the early signaling cascade of events described in \cite{conzelmann2008}. 
This model focuses on signaling from initial receptor binding (either of EGF 
or Ins), until the recruitment of the adaptor protein Grb2 in complex with Sos (a guanine nucleotide exchange factor that activates Ras). 
Grb2 is known as an adaptor protein
because of its ability to link EGFR activation to the activation of Ras and the downstream protein kinase cascade (e.g., RAF, MEK and ERK).
The model involves only eight proteins, which may combine into 2,768 different 
molecular species.
The interactions among these species are captured by a set of 42,956 reactions.

The reactions were translated into a rule-based model with $38$ reversible rules,
shown in Figure 7.
Eight node types arise:
\begin{displaymath} 
{\mathcal A}=\{EGF, EGFR, IR, Sos, Grb, IRS, Ins, Shc\}
\end{displaymath}
The contact map of the model is given in Figure 8a.
Each of the eight proteins is assigned a set of sites. 
For example, the representation of the receptor, $EGFR$, is assigned the set of sites
$\{a,b,c,d\}$.
The shaded sites in the figure are taken to have an internal state value. 
For example, site $b$ in $EGFR$ is allowed to have either of two internal state values - 
$\{\mathsf{u},\mathsf{p}\}$,
where $b_\mathsf{p}$ denotes that the site is phosphorylated. 
It is worth noticing that some sites have multiple binding partners, 
which denotes a competition (concurrency) for binding, 
because only one bond can be established at a time.
For example, site $a$ in $Grb$ has three possible binding partners.
Moreover, a self-loop at the site $d$ in $EGFR$ means that it can be bound to the site $d$ of another $EGFR$.
Therefore, one or two nodes of type $EGFR$ can be found in a single species.
Two major pathways are involved: one starting with the EGF receptor, $EGFR$, and another, starting at the insulin receptor, $IR$.
The two pathways share proteins.

By applying the algorithm of Figure 6 to the model, we obtain a reduction from a dimension of 2,768 species to 609 fragments.
The annotated contact map is given in Figure 8b. 
The interface of $Grb$ is split into two annotation classes, 
because no rule tests both sites $a$ and $b$ in $Grb$. 
Thus, the partition of the set of sites assigned to $Grb$ is $\{\{a\},\{b\}\}$, and it defines a set of fragments for which the reduction is exact.
Two fragment-based equivalent mixtures are shown in Figure 8c.
The largest species for this contact map counts 16 nodes (containing two $EGFR$ nodes, two $EGF$ nodes, four $Grb$ nodes, four $Shc$ nodes), while the equivalent fragment counts 12 nodes.

\section{Notes} 

\begin{enumerate}

\item The procedure for obtaining fragments which guarantee lumpability relations 
 correlates any two sites which are related directly or indirectly within a left-hand-side or a right-hand-side of a rule, and it hence enforces a strong independence notion between the uncorrelated sites. In turn, precisely such strong independence brings a possibility to effectively reconstruct the transient semantics of the original system. Despite such strong correlation notion, it was shown that the reduction can be significant, as shown over the EGFR/insulin crosstalk case study.
However, in most other test examples, the algorithm of Figure 6 reported the annotation equal to the species-based description. Indeed, a typical signaling cascade module involves a cascade of tests over pairs of sites, which are finally all correlated due to transitivity of annotation relation. In such a case, it is necessary to use a framework for \emph{approximate} reductions in order to quantitatively study coarse-grained executions. The approximate reduction proposed in \cite{approx} proposes the computation of error bound, while relying on knowing the generator matrix and transient distribution of the original process. To this end, the efficient numerical estimation of the error bounds is a compelling question for future work. 
Moreover, as ODE fragments are typically fewer than stochastic ones (for example, the presented EGF/insulin case study, the ODE fragments count $39$ and stochastic fragments $609$), it motivates to study whether ODE fragments can be used for exact simulation of stochastic traces, or, for correct computation of the transient distribution. 
To this end, the result of Kurtz \cite{kurtztg_1971_1} -- that the ODE model is a thermodynamical limit of the stochastic model -- is an important insight.

\item It is important to mention that the framework for reduction with fragments deals with providing more efficient executions of a given rule-based model (taken as the `ground truth'), while we do not address the problem of collecting the modelling hypothesis or validating that model with respect to experimental data. As a good model needs to be consistent with the observation, but also to predict behaviours which can be tested by observation, one immediate question is how to tailor the reduction to the high-level, qualitative experimental observation (for example, formation of a species, bimodality or causal relation between events). For example, for studying phenotypic variety, it sometimes suffices to use a model where each site is correlated only to itself  \cite{deeds_drift}.

\end{enumerate}

\section*{Acknowledgement}
This work has been supported by the SNSF Advanced Postdoc.\;Mobility Fellowship, grant number P300P2\_161067.

\clearpage

\bibliographystyle{spbasicunsorted}
\bibliography{Bibliography_Ch14_Petrov}

\clearpage


\section*{Figure captions}

\noindent \textbf{Figure 1}. An executable model captures some mechanistic understanding of how the systems under study work. Executing the models under various conditions can identify disagreements between hypothesized mechanisms and the experimental observations. This in turn provides new hypotheses or new experiments. This figure is reproduced with permission from Ref. \cite{FH07}. \\

\noindent \textbf{Figure 2}. Rule-set for Example 1. \\

\noindent \textbf{Figure 3}. Markov graph for $\xstov_0\equiv\{A,3B,C\}$. \\

\noindent \textbf{Figure 4}. Deterministic and stochastic models for Example 1. 
a) For volume $\volume=20\vunit$, the solution $\xdetv(t)$ of the deterministic model with initial state $\xdetv(0)=(1,3,1,0,0,0)\vunit$,  
and one scaled trajectory of a stochastic simulation $\xstov^{(\frac{\volume}{\vunit})}(t)$, for initial state $\xstov(0)=(20,60,20,0,0,0)$ (number of molecules). 
Rate values are set to $\drate_{1}=1 \vunit^{-1}s^{-1}$, $\drate_{2}=0.2 \vunit^{-1}s^{-1}$, $\drate_{1^-}=2 \vunit^{-1}s^{-1}$, $\drate_{2^-}=0.3 \vunit^{-1}s^{-1}$ and $\srate{1}=1 s^{-1}(\frac{\volume}{\vunit})^{-1}$ $\srate{2}=0.2 s^{-1}(\frac{\volume}{\vunit})^{-1}$, 
$\srate{1^-}=2 s^{-1}$, $\srate{2^-}=0.3 s^{-1}$. 
b) We integrated the CME for two initial states: $\xstov_1(0)=(1,3,1,0,0,0)$ (five equations of the model presented in Figure 3) and $\xstov_2(0)=(20,60,20,0,0,0)$ (set of 3,113 equations).
The three plots represent:
(solid lines) the solution $\xdetv(t)$ of the deterministic model with initial state $\xdetv(0)=(1,3,1,0,0,0)\vunit$,
(dashed lines) the scaled mean population for each species, for initial state $\xstov_1(0)$, that is,
$\frac{1}{3}\EE[\Xstov_1(t)]$ and,
(dotted lines) the scaled mean population for each species, for initial state $\xstov_2(0)$, that is, $\frac{1}{20}\EE[\Xstov_2(t)]$.\\

\noindent \textbf{Figure 5}. Stochastic fragments: motivating example. a) The Markov graph,for $\xstov_0\equiv\{\spec{A},3\spec{B},\spec{C}\}$;
b) The fragment-based Markov graph. \\

\noindent \textbf{Figure 6}. Algorithm for annotating the agent/molecule types of a rule-based program. \\

\noindent \textbf{Figure 7}. The set of rules for the EGF/insulin signaling crosstalk model.
The underlying mechanistic model is taken from \cite{conzelmann2008}.
The original model contains 42,956 reaction and 2,768 species.
Kappa syntax supports two types of shorthand notation: 
a site which simultaneously bears an internal state and serves as a binding site (for example, site $b$ of node type $EGFR$), and 
the dash symbol which denotes that the site is bound - for example, in Rule r10, $EGFR(b_u,d^-)$ denotes that site $d$ is bound.  \\

\noindent \textbf{Figure 8}. EGF/insulin signaling crosstalk model.
a) Contact map - summary of agent types and their interfaces (sets of sites). 
The gray-shaded sites bear internal value. 
b) Contact map annotation - summary of correlations between sites which must be preserved with fragments. 
c) Two reaction mixtures which are equivalent with respect to the annotation.
The green color denotes phosphorylated state.
d) An example of a Kappa rule and the site-graph rewrite rule:
 $EGFR(b_{\mathsf{u}},d)$ denotes a site-graph 
  with one node of type $EGFR$ and interface $\{b,d\}$ and internal evaluation of site $b$ to $\mathsf{u}$.




\clearpage

\begin{figure}[h]
\begin{center}
\includegraphics[width=1\linewidth]{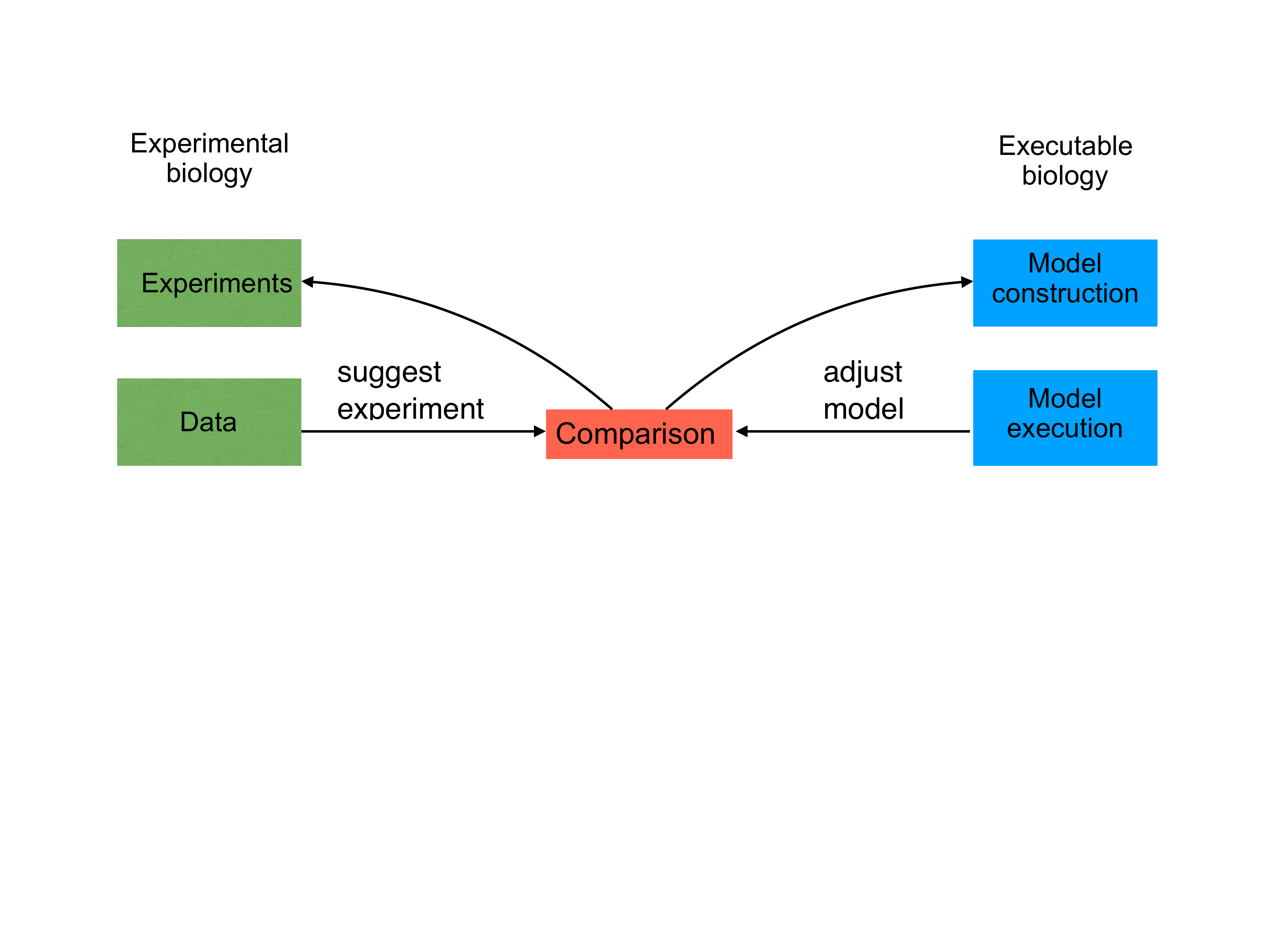}
\caption{}
\label{fig:1}
\end{center}
\end{figure}

\clearpage

\begin{figure}[h]
\begin{center}
  {\includegraphics[scale=0.4]{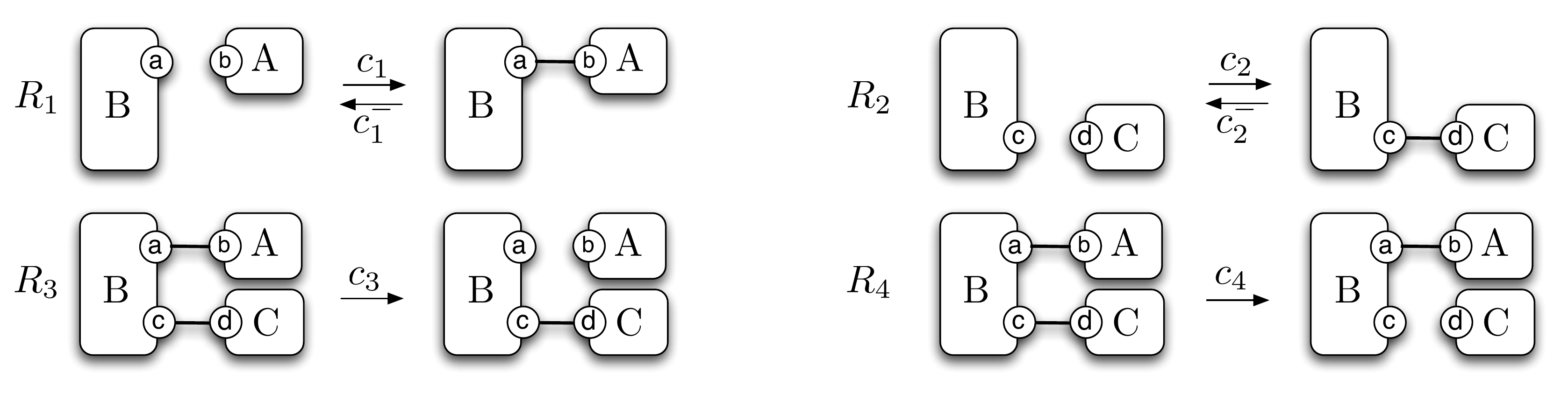}}
\end{center}
\caption{}
\label{fig:2}
\end{figure}

\clearpage

\begin{figure}[ht]
\begin{center}
  {\includegraphics[scale=0.4]{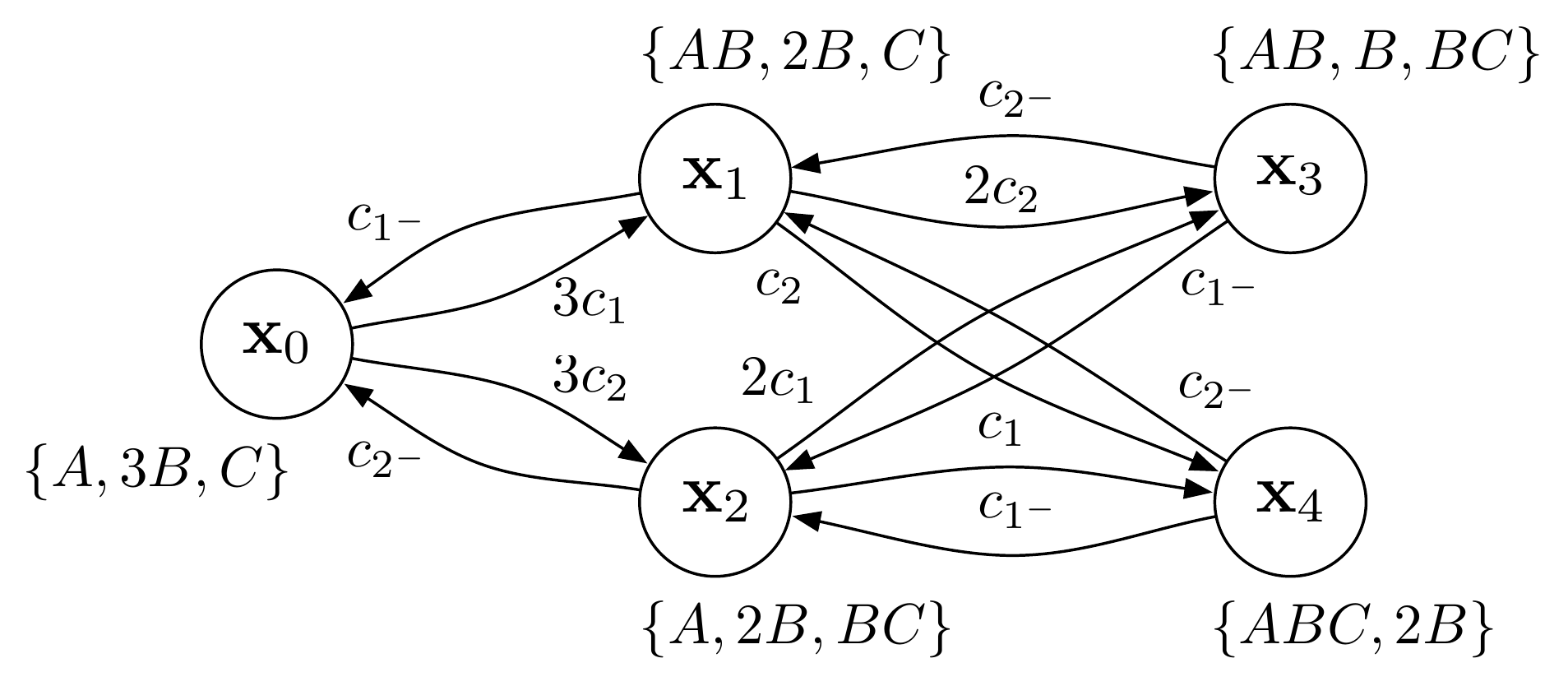}}
\end{center}
\caption{}
\label{fig:3}
\end{figure}

\clearpage


\begin{figure}[t]
\begin{center}
\includegraphics[width=1.\linewidth]{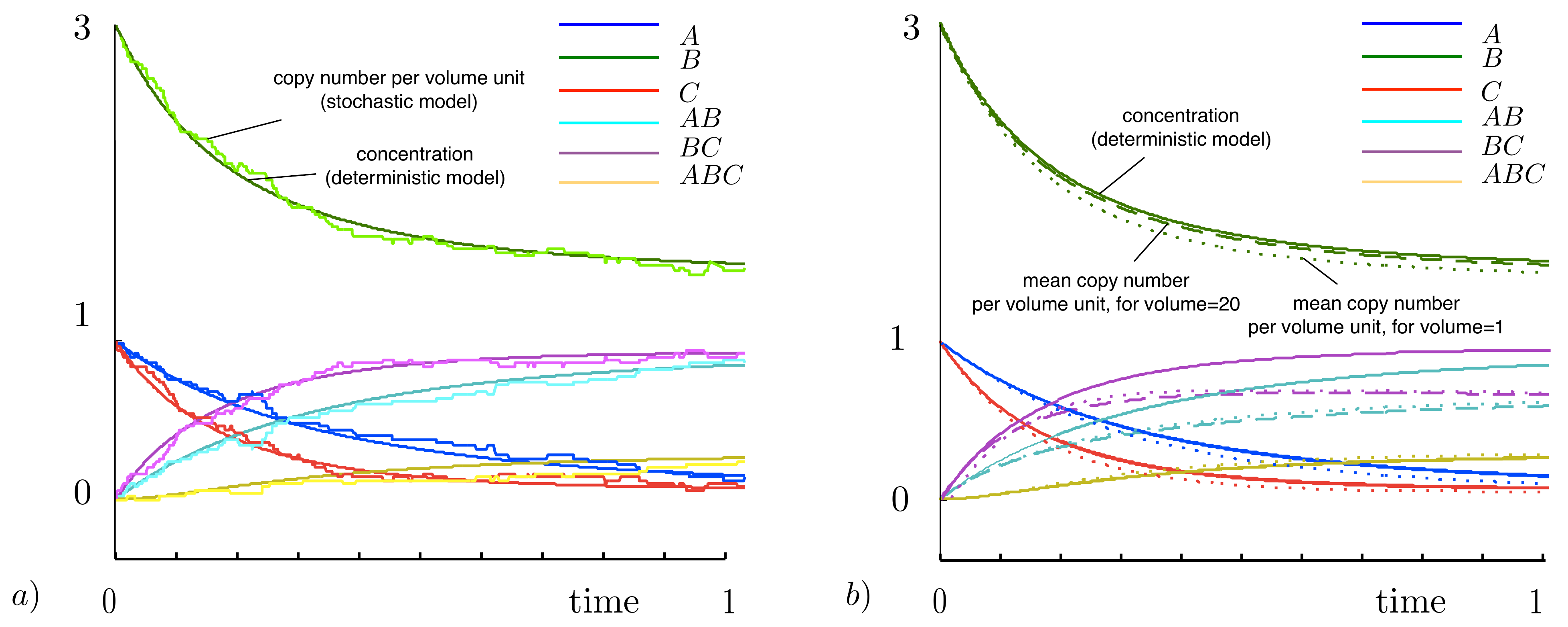}
\end{center}
\caption{}
\label{fig:4}
\end{figure}

\clearpage

\begin{figure}[t]
\begin{center}
  {\includegraphics[scale=0.4]{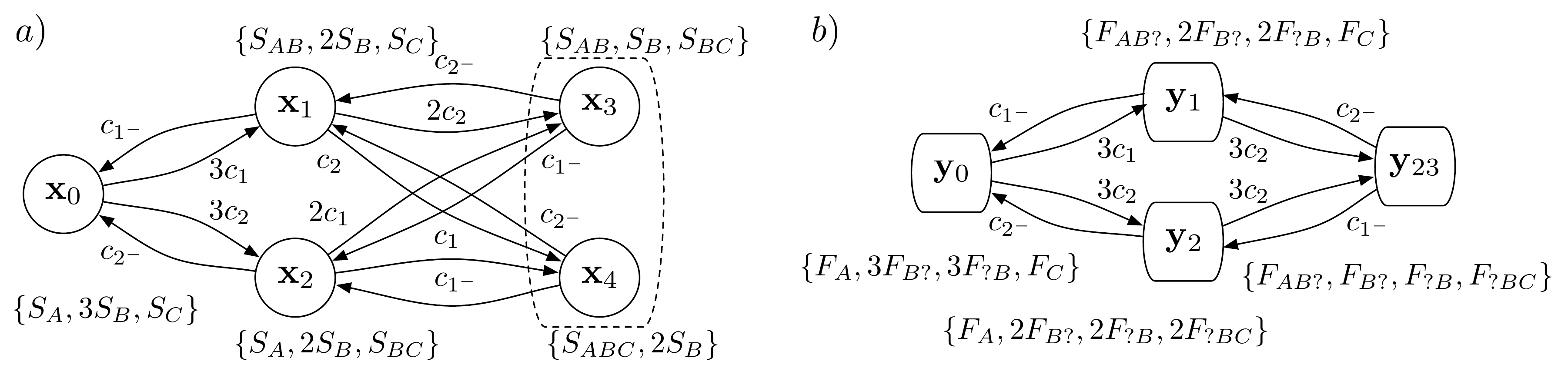}}
\end{center}
\caption{}
\label{fig:5}
\end{figure}

\clearpage

\begin{figure}[h]
\begin{center}
  {\includegraphics[scale=1.0]{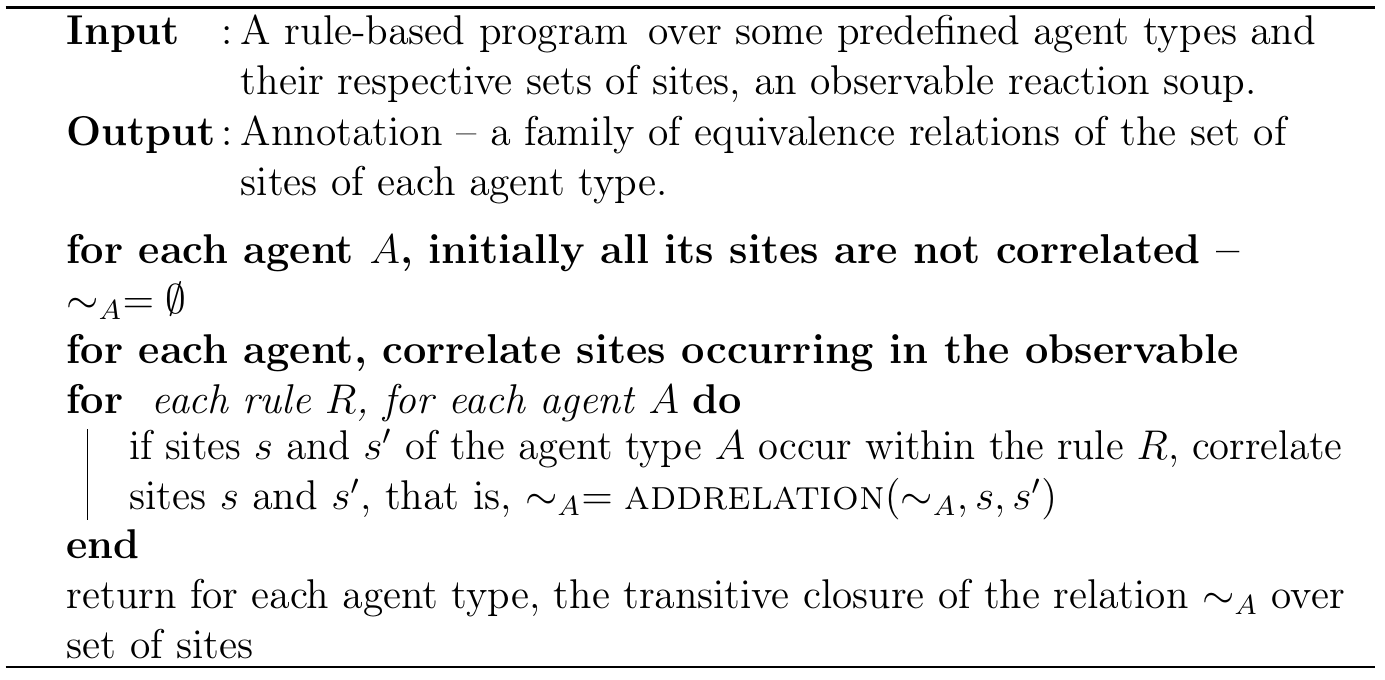}}
\end{center}
\caption{}
\label{fig:6}
\end{figure}


\clearpage

\begin{figure}[t]
\begin{center}
\includegraphics[scale=0.35]{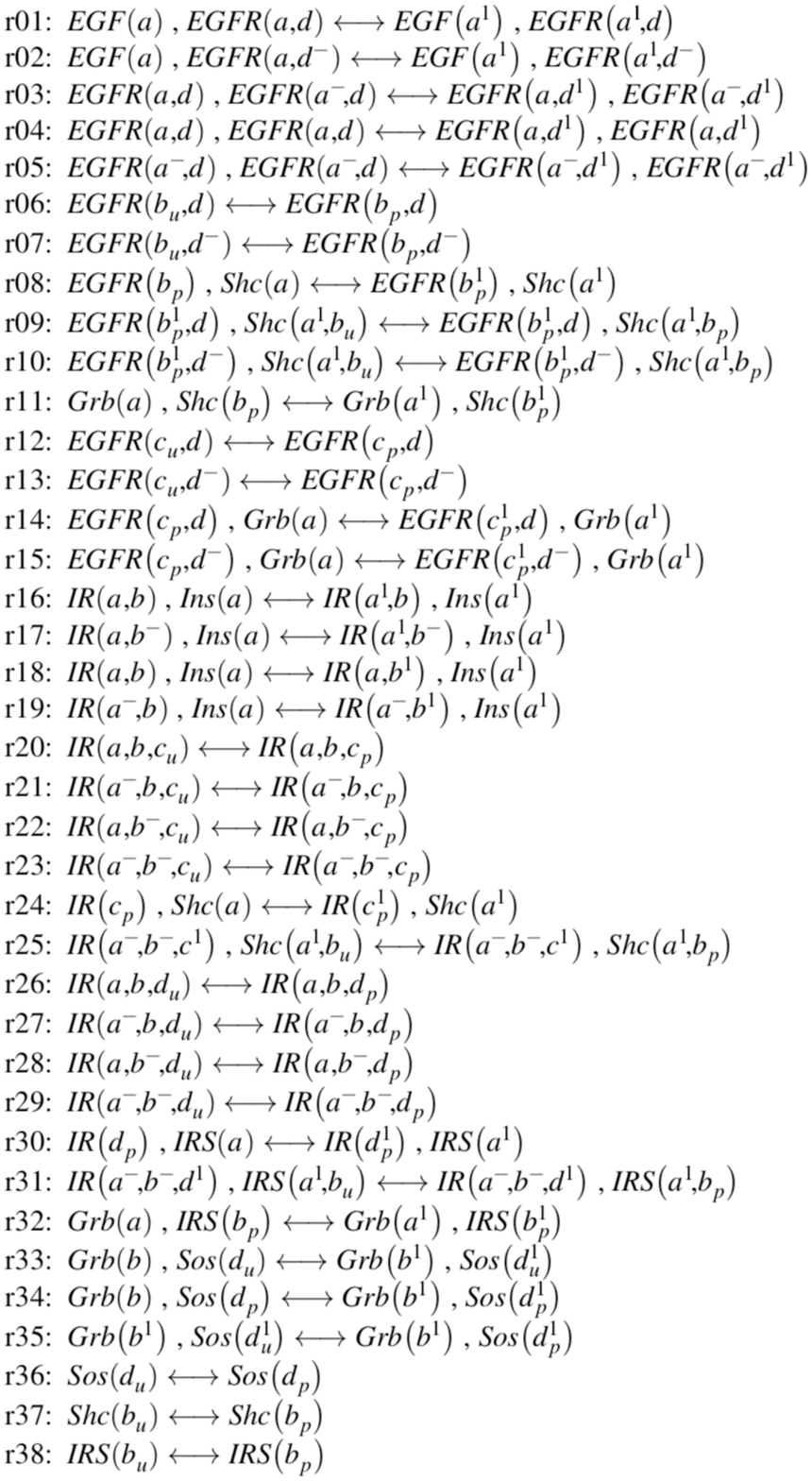}
\end{center}
\caption{}
\label{fig:7}
\end{figure}

\clearpage

\begin{figure}[t]
\begin{center}
  \includegraphics[width=1\linewidth]{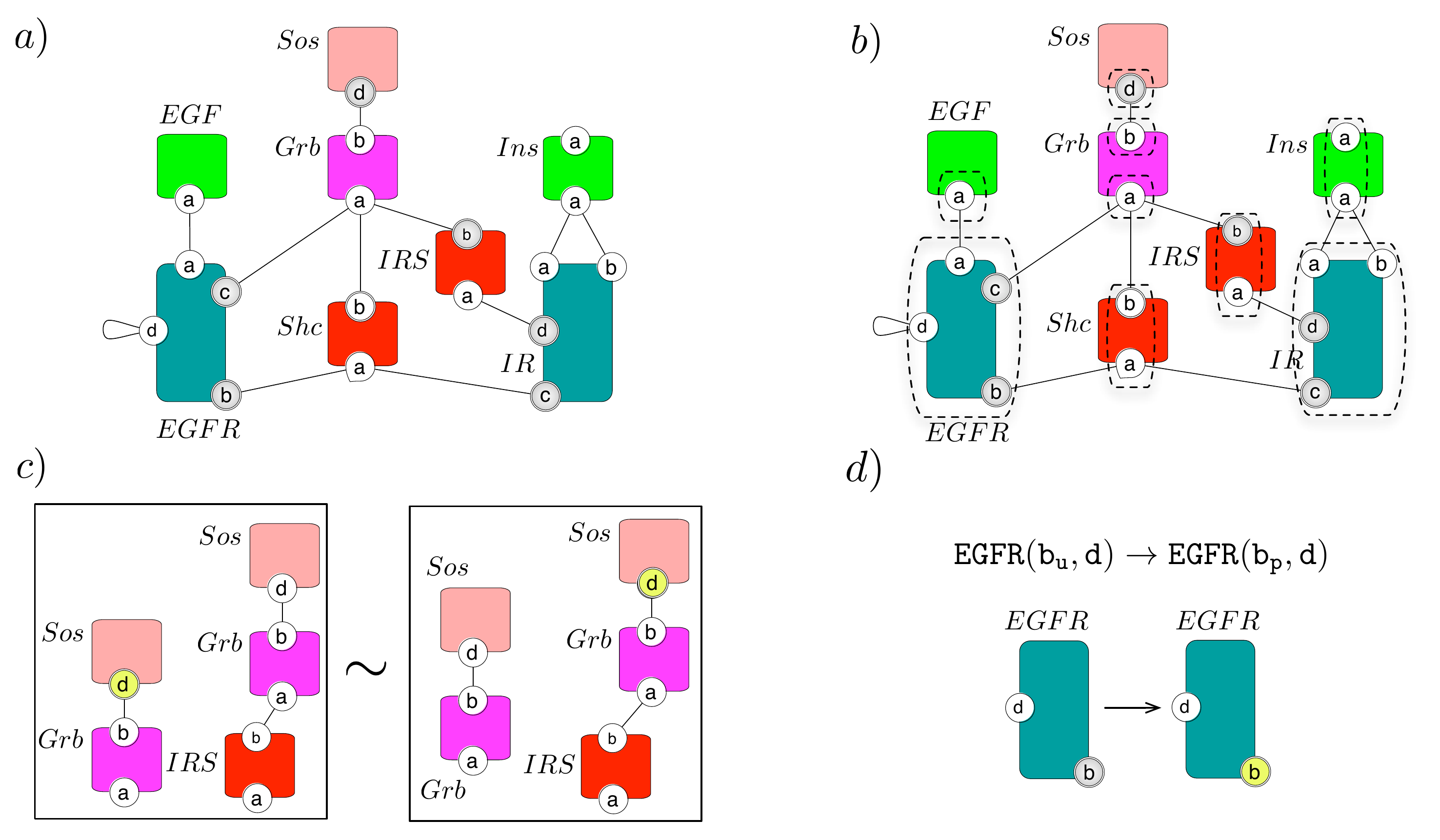}
\end{center}
\caption{}
\label{fig:8}
\end{figure}

\end{document}